# A COMPARISON BETWEEN CODYRUN AND TRNSYS, SIMULATION MODELS FOR THERMAL BUILDINGS BEHAVIOUR.

F. LUCAS, T. MARA, F. GARDE, H. BOYER

Chercheurs au Laboratoire de Génie Industriel, Université de la Réunion,

15, av. René Cassin 97703 Saint Denis Réunion.

Tel : (0 262 ) 93 82 16 ; Fax : (0 262) 93 81 66;

E.Mail : lucas@iremia.univ-reunion.fr

ABSTRACT:

Simulation codes of thermal behaviour could significantly improve housing construction design. Among the existing software, CODYRUN and TRNSYS are calculations codes of different conceptions. CODYRUN is exclusively dedicated to housing thermal behaviour, whereas TRNSYS is more generally used on any thermal system. The purpose of this article is to compare these two instruments in two different conditions . We will first modelize a mono-zone test cell, and analyse the results by means of signal treatment methods. Then, we will modelize a real case of multi-zone housing, representative of housing in wet tropical climates. We could so evaluate influences of meteorological and building description data on model errors.

## I. INTRODUCTION

Comfort demands and reduction of air conditioning unit prices lays the problem of energetic supplies in French Overseas Departments. Indeed, air conditioning unit sales have considerably increased during the last few years. Their simultaneous use, at the warm hours of the days generates high consumption peaks which must be absorbed growing needs of supply, for a high cost. In order to this trend, it becomes essential to build comfortable housings in a thermal and economical viewpoint. Our purpose is to reduce the use of air-conditioning for the benefit of natural ventilation and solar protection of housing. As natural ventilation becomes insufficient to ensure occupant comfort, it is necessary to have recourse to mechanical ventilation [1] or air conditioning. In this case, equipment consumption must be diminished.

The simulation of thermal behaviour allows to propose improvements concerning housing conceptions. More than hundred building thermal simulation models have been developed during the last decade. Among them CODYRUN is a multi-model calculation code dedicated to thermal simulations of mono-zone or multi-zone housings and for various types of climate. This software has already been used for the validation of prescriptions concerning low energy cooling of buildings in tropical climates [2]. The standard models of housing developed with TRNSYS allow to study mono-zone or multi-zone cases with type 56 « multi zone building ».
These two softwares have different configurations and were build for different purposes. Simulations have been carried out on two different examples (mono-zone and multi-zone buildings) and compared to experimental results. Mono-zone simulations have been done upon an adiabatic test cell, whose thermal and physical characteristics are well-known. The multi-zone building which has been chosen, is an individual house, typical from humid tropical regions.

## II. PRESENTATION OF CODYRUN

CODYRUN is a dedicated software tool usable by researchers and suitable for professionals. More precisely, CODYRUN is a multizone software integrating both natural ventilation and moisture transfers, developed on a PC micro computer [3,4]. One of its most interesting feature is to offer the expert thermician a wide range of choices between different heat transfer models and meteorological reconstitution parameter models. Because of the amount of information necessary to describe one building, CODYRUN takes advantages of a user friendly front-end, based on Microsoft Windows environment.

The lumped capacities analysis is a powerful method of investigation in the thermal analysis of systems. Considering the thermal behaviour of a building, its thermal state is determined by the continuous field of temperatures, concerning all points included in the physical limits of the building. The constitution of a reduced model with a finite number of temperatures, is possible by assuming few assumptions for simplifications. As a result, we consider mono-dimensional heat conduction in walls and well mixed air volumes of each thermal zone. The obtained equations are solved using a classical finite difference scheme.

This study focuses on thermal aspect as no airflow is taken into account at this step.

## III. PRESENTATION OF THE TRNSYS MODEL

TRNSYS is a widespread simulation environment whose aim is to simulate the transient behaviour of thermal systems [5]. It uses a modular approach in which the system components are described by FORTRAN subroutines. Projects modelizing mono-zone and multi-zone buildings lay on the use of type 56 « Multi-zone Building » and associated utilities PREDBID and BID [6]. TRNSYS and its utilities are encapsulated in IISIBAT interface which allows a fast input of simulation project [7].
In the Type 56, the heat conduction in walls, ceilings and floors are modeled using the ASHRAE transfer function approach. The short-wave and long-wave radiation transfers are calculated with an area ratio method.

## IV. . RESULTS AND DISCUSSION

### MONO-ZONE SIMULATION

EXPERIMENTAL DEVICE. Simulations have been done upon a fully instrumented experimental cell (fig.1). It is a 22,5 m3 enclosure. The walls are made of sandwich panels composed of two 7 mm layers of cement fibre boards with 6 cm of polyurethane foam. The roof is made of a sandwich panel with cement fibre board, polyurethane layer and aluminium sheet. The test cell lies on a concrete paving stone floor and a polyurethane layer. The experimental instruments measure weather solicitations and inner thermal conditions of the precinct.

A weather station located near the cell is aimed to get the climatic data (dry air temperature, relative humidity, direct radiation, diffuse radiation, wind speed, wind direction). These data allow to simulate the cell behaviour with CODYRUN and TRNSYS on identical weather solicitations. As we know quite well the building thermal and physical characteristics, we will evaluate the influence of weather parameters on the two models accuracy.

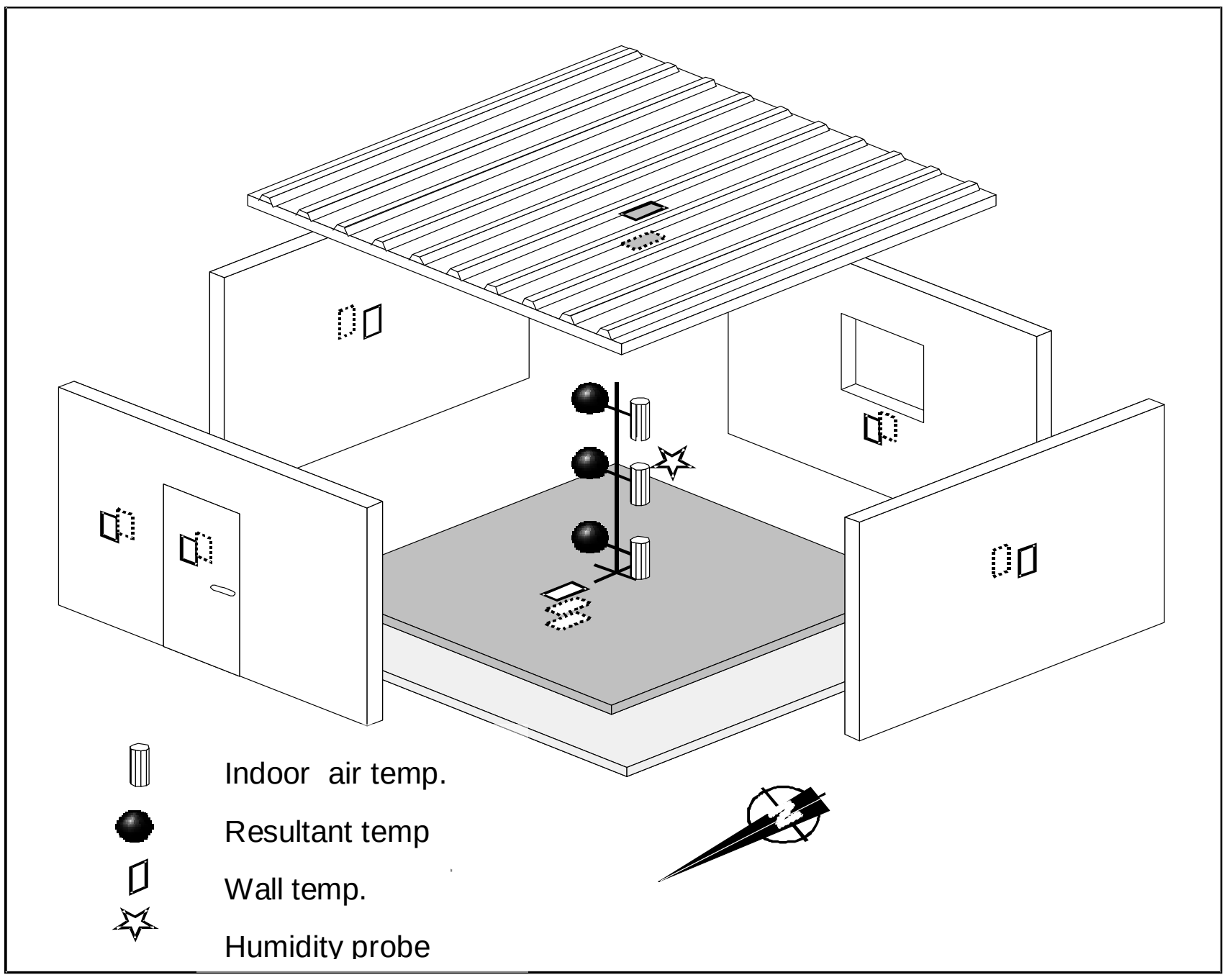

*fig 1: Mono-zone test cell and associated sensors.*

RESULTS: In the mono-zone case, the visual analysis of the simulation results shows that both codes give good results. Indeed, the residual averages are well centred and the standard deviation remains low.(Tab.1)

| Residuals (Experiment – model) | CODYRUN | TRNSYS |
|---|---|---|
| Mean | 0.088 | - 0.191. |
| Standard deviation | 0.44 | 0.43 |

*Tab 1: Mean and standard deviation of residuals for mono-zone simulations*

Analysis of the power spectrum density of the indoor dry air temperature residuals (Fig 2) shows the residuals isn't a white noise since it has a period of 24h (frequency=0.0417 $h^{-1}$). This means that the modelisation of the test cell is not accurate neither with TRNSYS nor with CODYRUN [8]. We can also notice that the power spectrum density of the residuals spreads principally between 0 and 0.1 $h^{-1}$ where the power is superior to 0.1°C². [9]

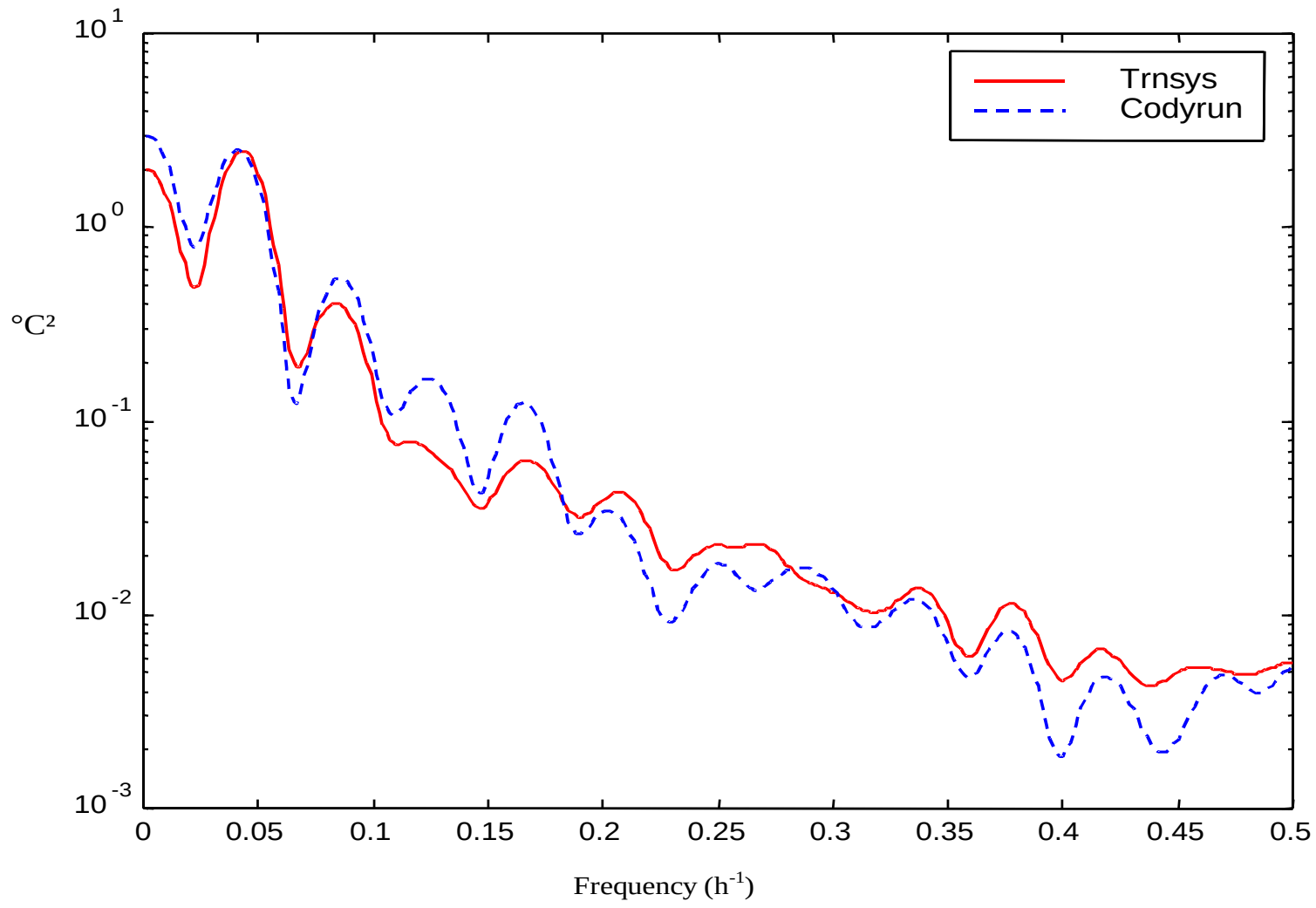

*Fig 2: Power spectrum density of indoor temperatures residuals for CODYRUN and TRNSYS*

To see which weather data (or excitations) are responsible of the simulation error we will first analyse the squared coherency spectrum and then calculate the contribution from each excitation to the total variance of the residuals .

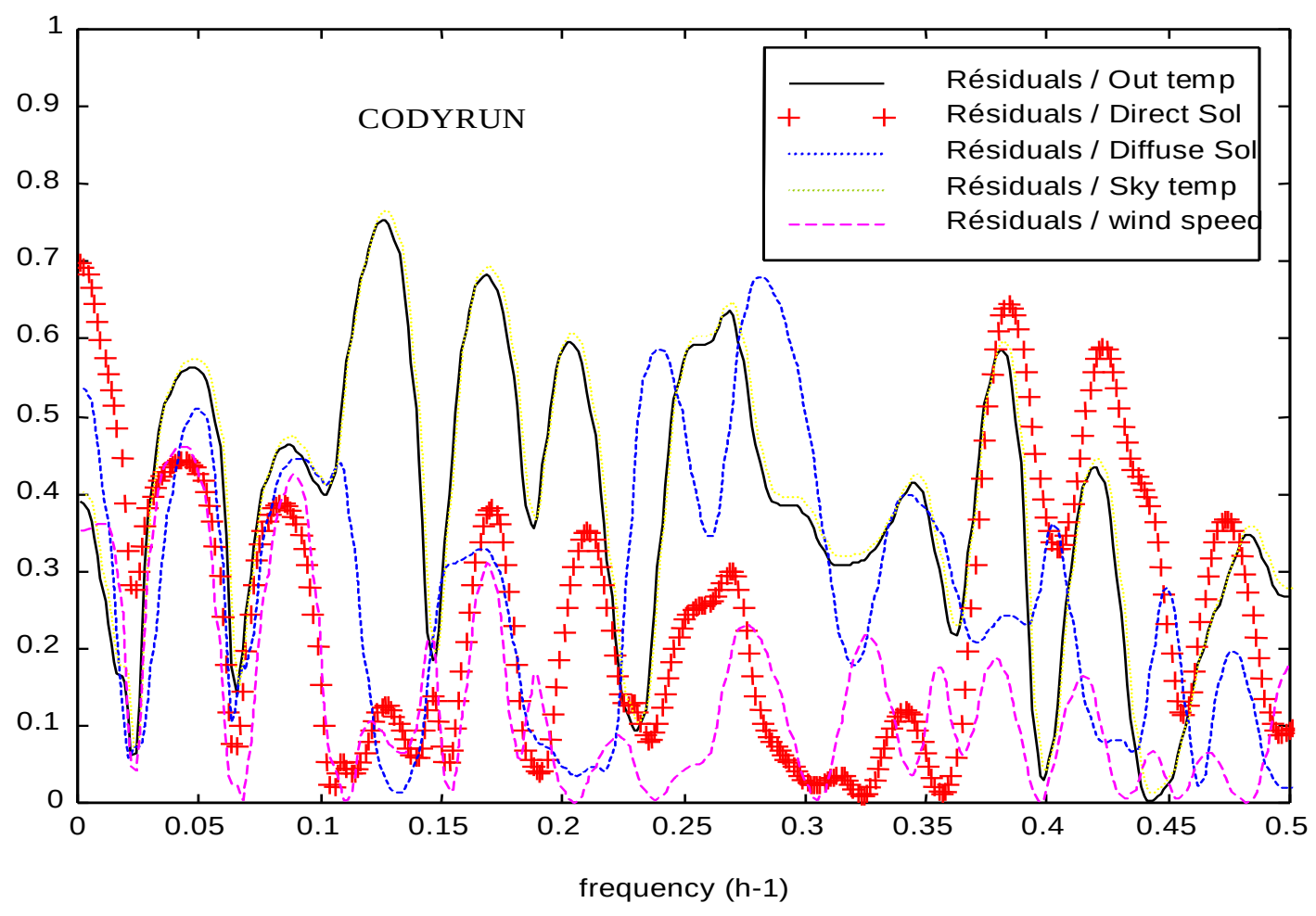

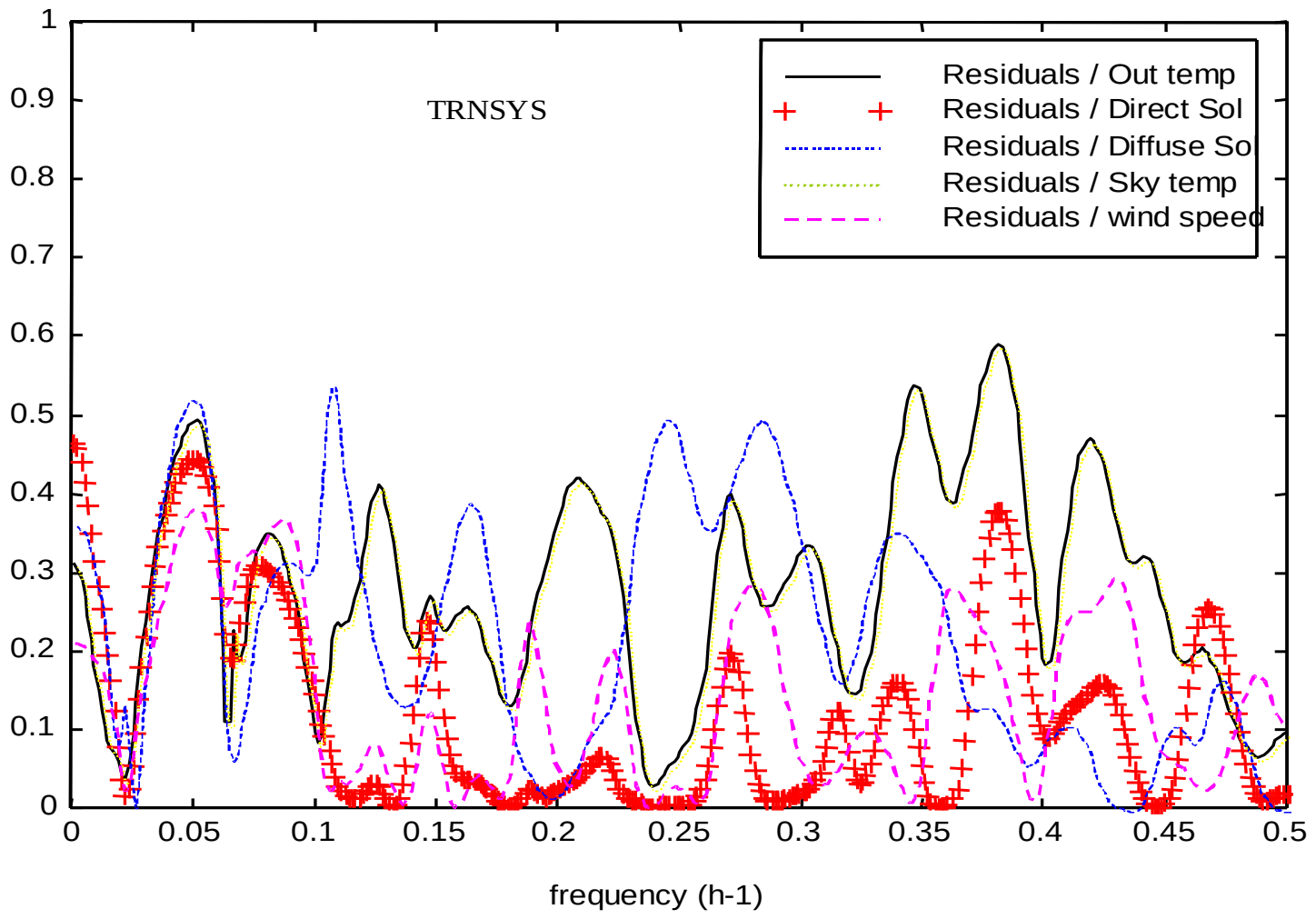

*Figure 3: Squared coherency spectrum between residuals for CODYRUN and TRNSYS*

The squared coherency spectrum allows to measure the linear correlation between two signals. Figure 3 shows the squared coherency spectrum between the residuals (simulation errors) and the weather data (excitations of the test cell) for TRNSYS and CODYRUN. First, we have to notice that the sky temperature is calculated from the outdoor air temperature. Therefore the squared coherency spectrums with the sky temperature and outdoor air temperature are the same.

We can see that the correlation is stronger with CODYRUN (since the squared coherency spectrum is between 0 and 0.8) than with TRNSYS (0 to 0.6). A small squared coherency spectrum means that the excitations considered are less correlated with the residuals or that the relations between the excitations and the residuals are not linear. We can also notice that for TRNSYS simulation errors, the correlation is stronger first with the diffuse solar (frequency between 0.02 and 0.32 $h^{-1}$) and then with the out air temperature (or sky temperature) and with the direct solar (frequency between 0 and 0.02 $h^{-1}$). Whereas for CODYRUN, correlation is stronger with the out air or sky temperature (0.02 and 0.32 $h^{-1}$) then with the direct solar (0 and 0.02 $h^{-1}$) and then with the diffuse solar.

We are now going to decompose the variance of the residuals to find the contribution from each excitation (we will consider out air temperature, direct solar and diffuse solar). But in order to account for any correlation between the excitations, these are sorted by their order of importance, then the part of each excitations which is correlated to the previous one is removed [9,10].

According to the analysis before, we will consider four frequency ranges:

[0;0.02] $h^{-1}$ low frequency

[0.02;0.08] $h^{-1}$ including the first harmonic

[0.08;0.32] $h^{-1}$ including the second and third harmonics

[0.32;0.5] $h^{-1}$ high frequency

Figure 4 shows the decomposition of the air temperature prediction errors for TRNSYS and CODYRUN. This shows, concerning TRNSYS, that the diffuse solar and the out air or sky temperature are the dominant sources of errors in the two first frequency ranges where lies the principal power of the residuals (89%) lies. Yet, the main part of the errors does not seem to be explained by the solicitations chosen It may be due to a non linear relationship between the weather data and the residuals. In contrast, concerning CODYRUN, the weather data seems to better explain the simulation errors as the unexplained part is low. We can notice that in the low frequency range [0;0.02] $h^{-1}$, direct solar seems to be the first responsible for the discrepancy. But in the second frequency range [0.02;0.08] $h^{-1}$, the simulation errors are likely to be due to the way the out air or sky temperature is taken into account by CODYRUN.

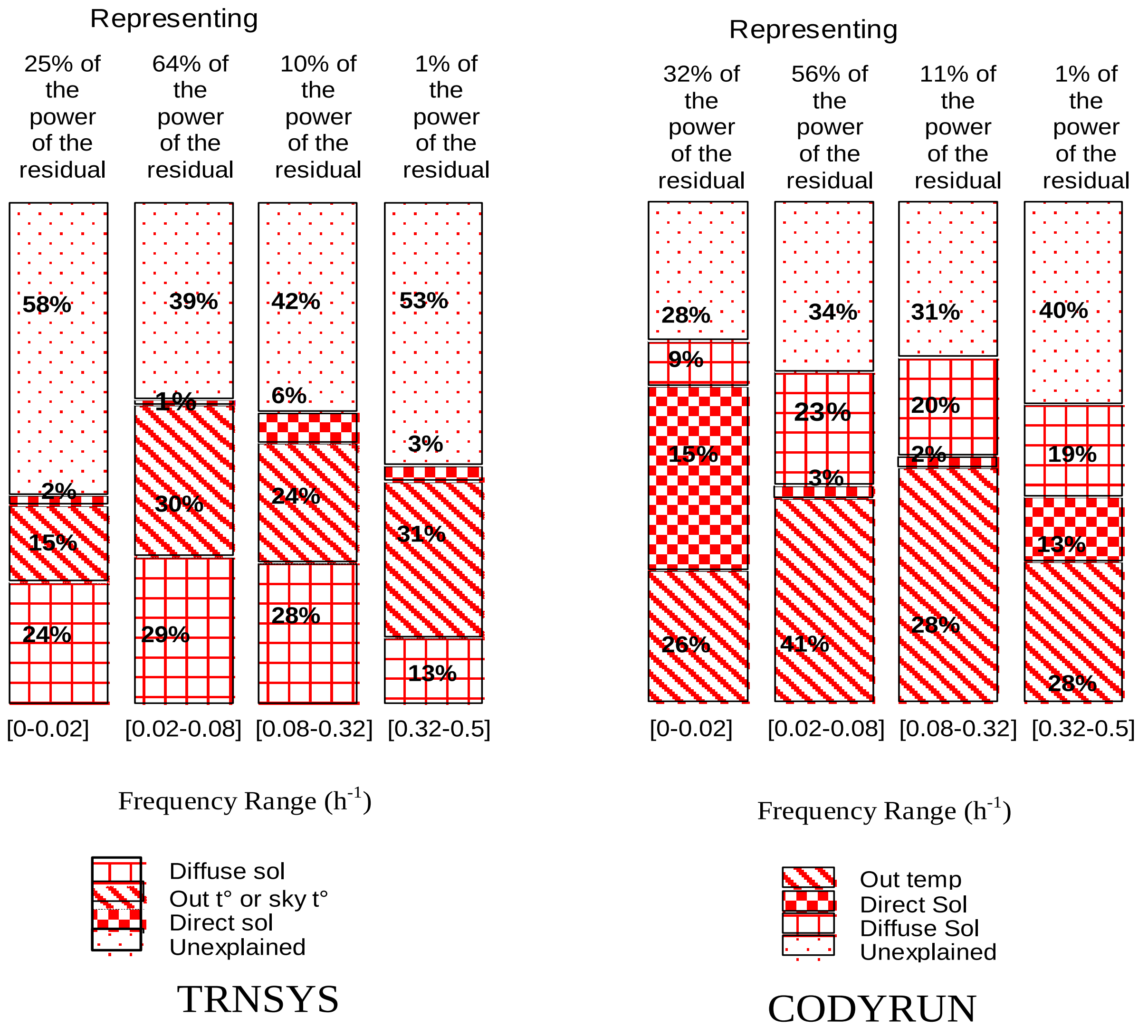

*Figure 4: decomposition of the air temperature prediction errors for TRNSYS and CODYRUN*

MULTI-ZONE SIMULATIONS

EXPERIMENTAL DISPOSAL. Simulations are based on a type of classical housing in humid tropical climate mainly composed with light walls. Exterior walls are made of a breeze block layer covered with cement rendering or wooden weather boarding. Inner walls are made of a air layer between two wood fibreboard layers. The roof is made of a steel sheet, a air layer and a fibreboard. Moreover, the house is surrounded by a veranda and closed, adjacent rooms that constitute solar protections. We will distinguish five thermal zones in this construction: a kitchen, a bathroom, a toilet and two bedrooms (Fig. 5)

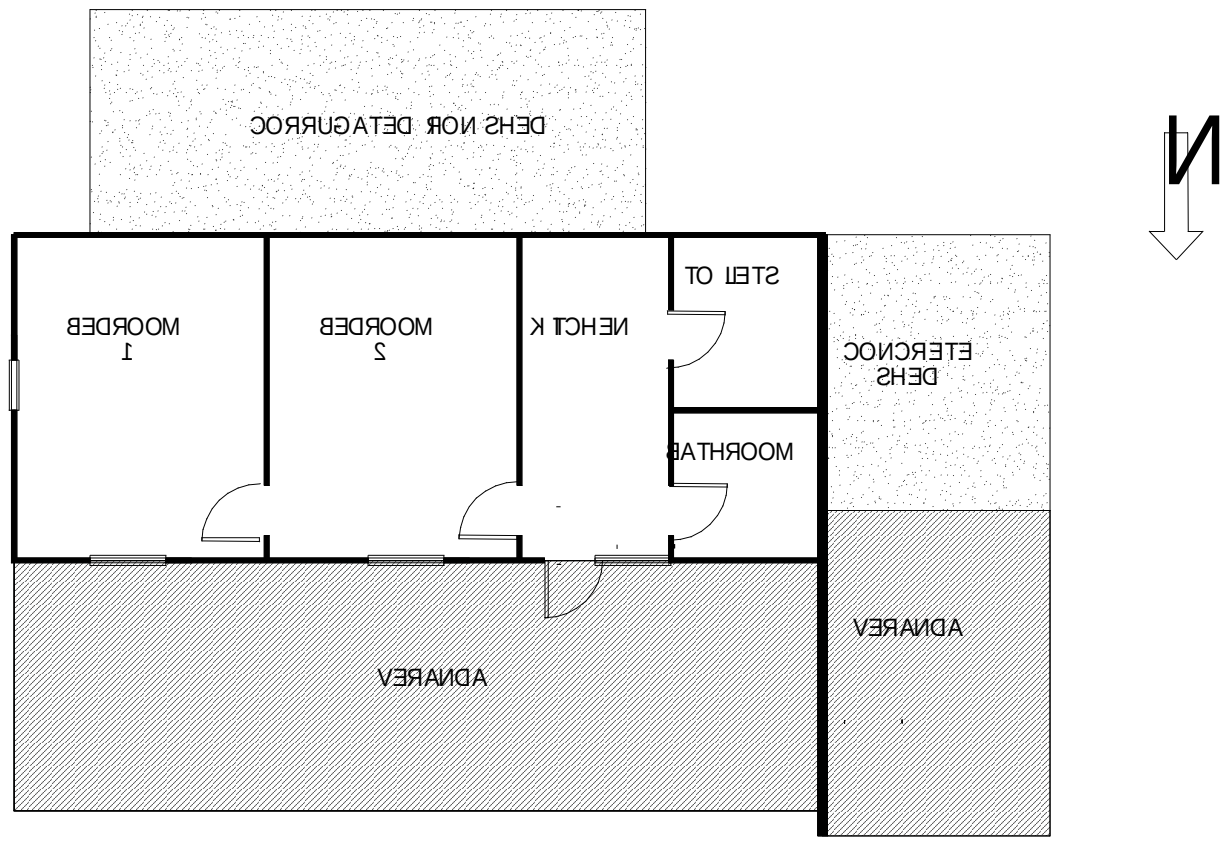

*Fig 5: Multi-zone building*

<u>RESULTS</u>. Considering the precision of dry air temperature measures (about 1°C), the values obtained in the different zones by simulation are in accordance with reality. Standard deviations as well as residual averages generally remain inferior to 1°C except for the toilet zone (Tab.2). However residuals are superior to those obtained with the mono-zone simulations. This can be explained because components of the cell walls are well-known whereas it's difficult to evaluate the right composition of walls in a real existing building.

| | Bedroom 1 | Bedroom 2 | Kitchen | Bathroom | Toilets | |
|---|---|---|---|---|---|---|
| Mean | - 0.76 | - 0.63 | - 1.02 | - 0.38 | - 0.88 | CODYRUN |
| | 0.43 | 0.72 | 0.19 | - 0.21 | 0.87 | TRNSYS |
| Standard | 0.52 | 0.51 | 0.39 | 0.40 | 1.44 | CODYRUN |
| deviation | 0.74 | 0.72 | 0.79 | 0.62 | 2.02 | TRNSYS |

*Tab 2: Means and standard deviations of residuals for multi-zone simulations*

Nonetheless, we will notice that CODYRUN tends to underestimate the zones temperatures. Indeed, the residual between experimental temperature and the value given by CODYRUN highly increases during the night. Exchanges with the sky vault become then predominant. During the day, results are better for solar radiation as it is preponderant. It is possible to explain this phenomenon by the importance of the sky temperature for light construction simulations in tropical climate. The model of sky temperature used (Tsky = Ta - 6) can be improved in order to better take in account the radiative transfers.

TRNSYS calculates conductive transfers throughout the walls by the transfer function method which needs particular precautions in the case of light construction walls. Indeed, it is not possible to calculate the transfer function coefficients by the utility BID for low mass walls. We have ,then ,to define a "massless" layer only characterised by its thermal resistance. Then the walls composed with fibreboard + air layer + fibreboard, doors, shutters, and above all the roof cannot be modelised with the help of transfer function coefficients.

Garde [9,10] has underlined the importance of long waves radiative transfers into air layer of the roof. We have calculated a thermal resistance similar to the air, in order to better consider the role of roofs with air layer. We have been using an equivalent thermal resistance of air: $R_{air}$ = 9.26 $W.m^{-1}.K^{-1}$ for thickness of 10 cm.

## V. CONCLUSION ET PERSPECTIVES

CODYRUN and TRNSYS have slightly different objectives and configurations, but the simulation results are comparable and in accordance with the experimental results. This inter-software comparison demonstrates that, in humid tropical climate, some parameters turn to be essential. Spectral analysis points out that diffuse radiation and outdoor temperature generate the most important part of the TRNSYS residual, whereas CODYRUN doesn't take the direct radiation and the outdoor temperature into account very well.

These simulations on a test cell as well as on a real housing have brought important information about thermal transfers modelisation with both codes. But the airflow between zones have not been taken into account. The typical housing of tropical climates is characterised by a great permeability. The next stage will consist in a comparison with air transfer in order to improve the housing thermal quality by encouraging natural ventilation .

## VI. APPENDIX

Mono-zone simulations: Temperatures and residuals

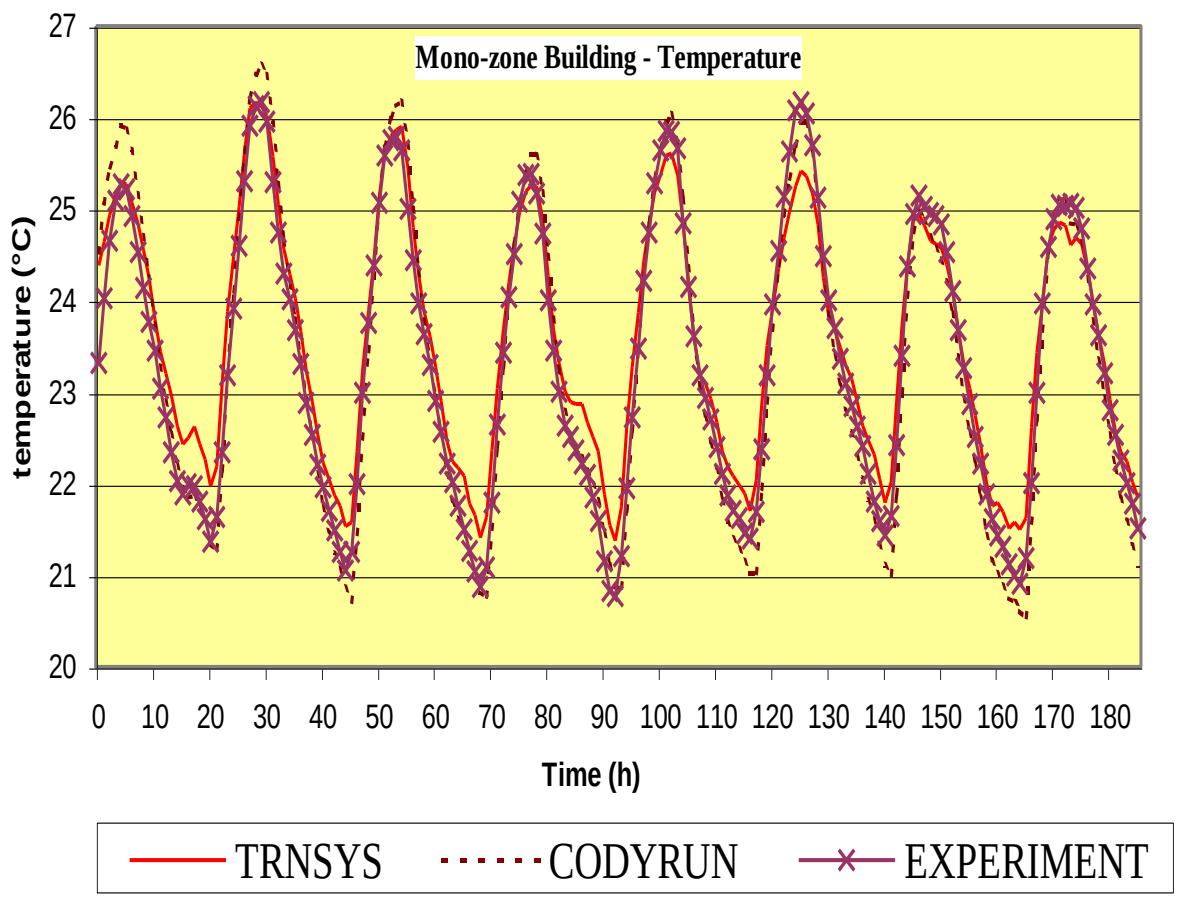

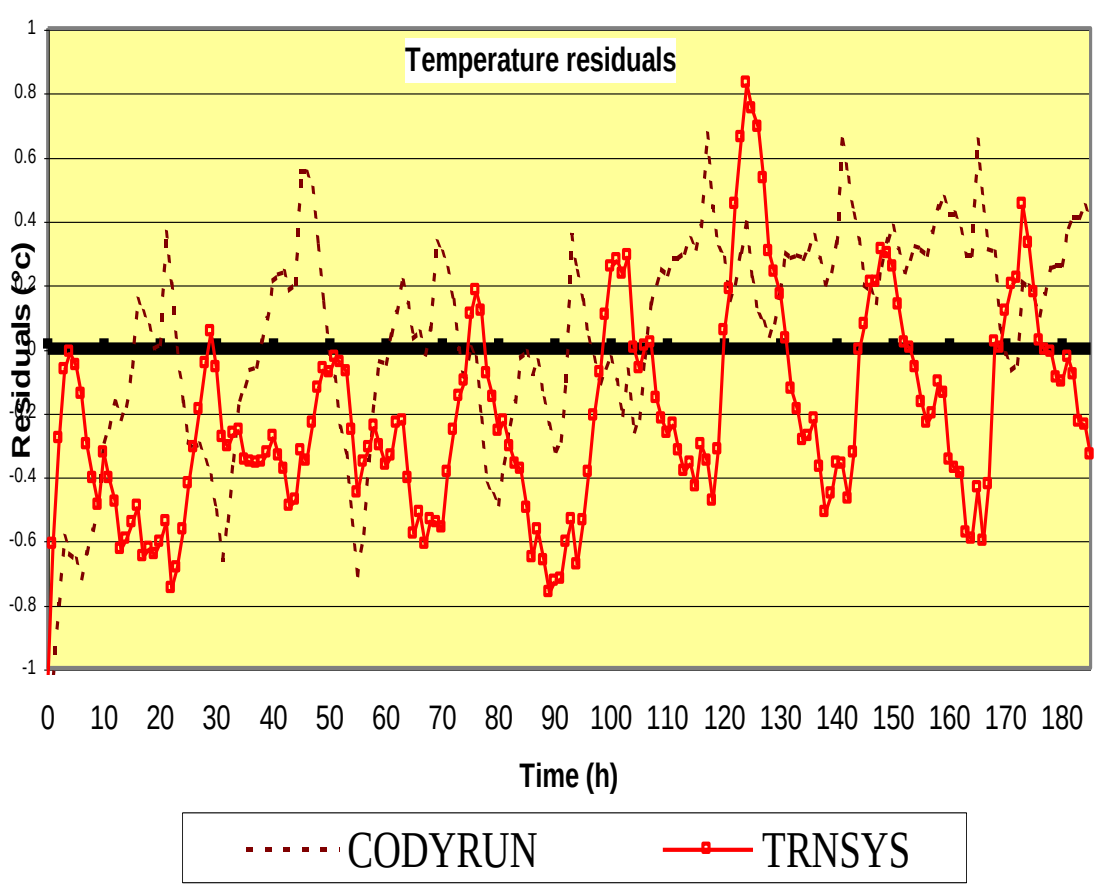

# Multi-zone simulations: Temperatures and residuals

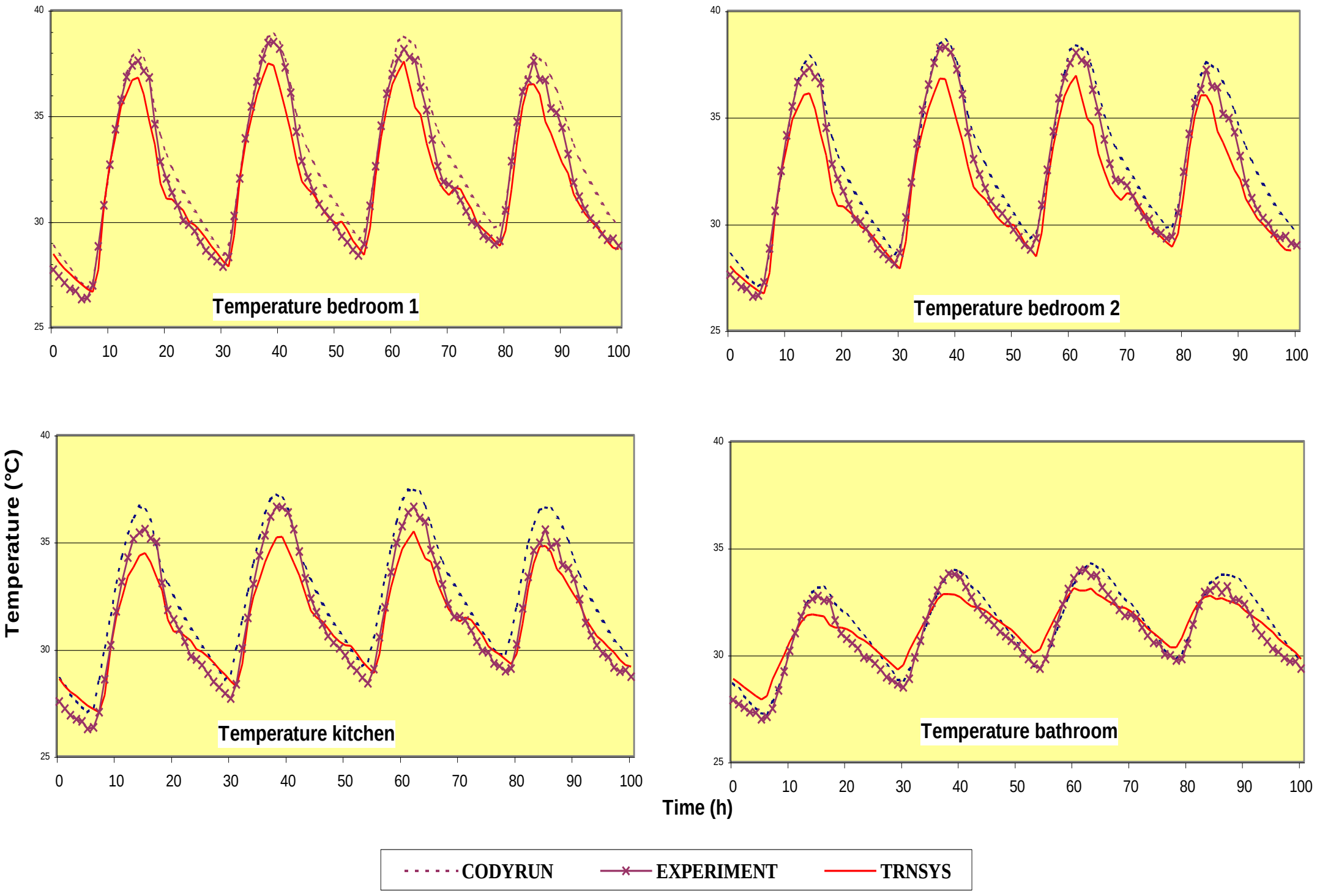

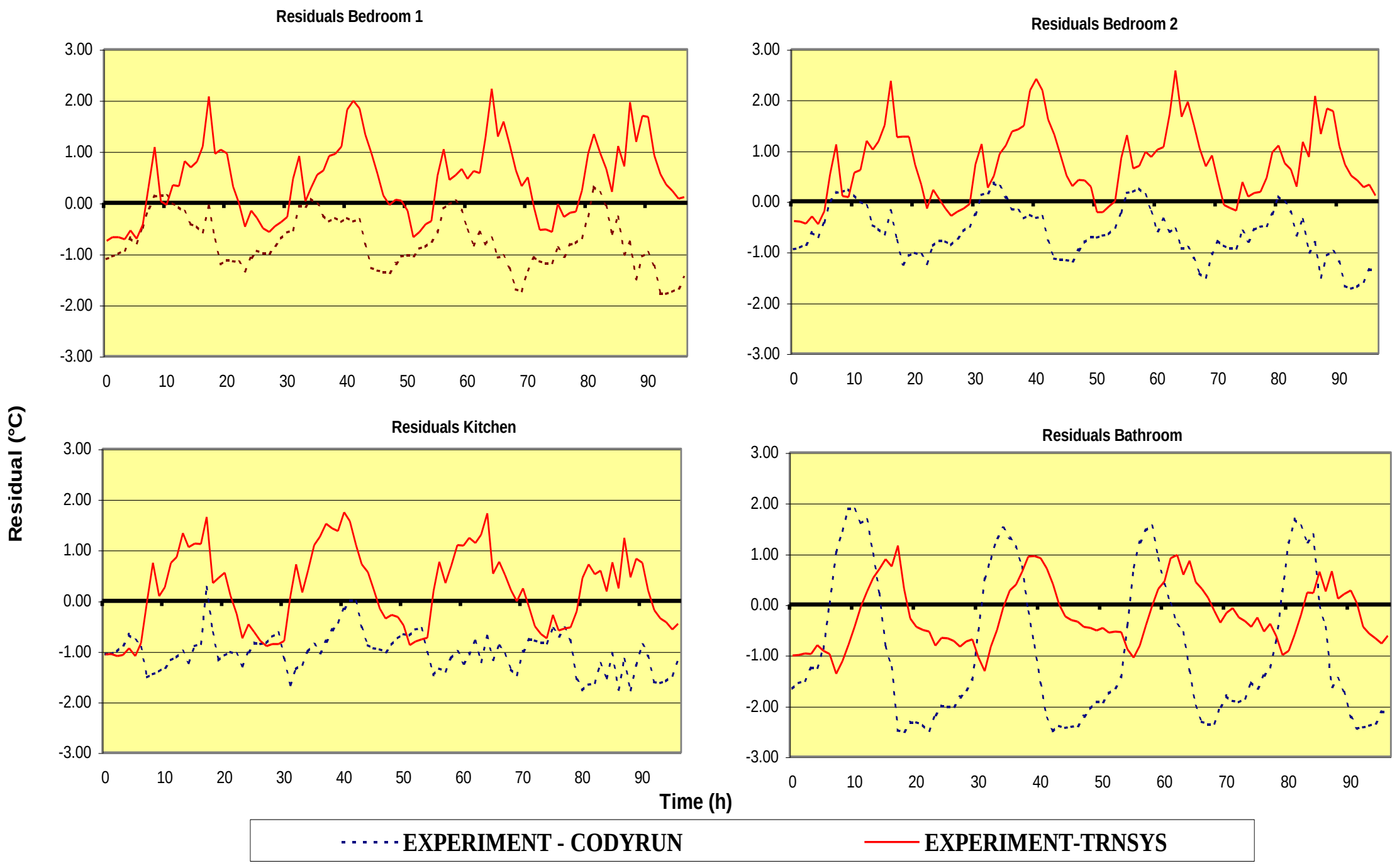